\documentstyle{article}
\makeatletter
\def\revtex@ver{3.0}
\def\revtex@date{10 Jan 93}
\def\revtex@org{AAS}
\def\revtex@jnl{AAS}
\def\revtex@genre{preprint}
\def\revtex@pageid{\xdef\@thefnmark{\null}
\@footnotetext{This \revtex@genre\space was prepared with the
		   \revtex@org\space \LaTeX\ macros v\revtex@ver.}}
\def\genre@MS{manuscript}
\def\genre@PP{preprint}
\ifx\revtex@genre\genre@PP
\ifnum\@ptsize<1
\fi
\fi
\def\ps@plaintop{\let\@mkboth\@gobbletwo
\def\@oddfoot{}\def\@oddhead{\rm\hfil--\space\thepage\space--\hfil}
\def\@evenfoot{}\let\@evenhead\@oddhead}
\ps@plaintop
\@rightskip=\z@ plus 4em\rightskip\@rightskip
\def\@tightleading{1.1}
\def\@doubleleading{1.6}
\def\baselinestretch{\@doubleleading}
\def\tighten{\def\baselinestretch{\@tightleading}}
\let\tightenlines=\tighten
\def\singlespace{\def\baselinestretch{\@tightleading}\normalsize}
\def\doublespace{\def\baselinestretch{\@doubleleading}\normalsize}
\def\sec@upcase#1{\relax{#1}}
\def\eqsecnum{
\@newctr{equation}[section]
\def\theequation{\hbox{\normalsize\arabic{section}-\arabic{equation}}}}
\tighten
\def\@journalname{The Astropolitical Journal}
\def\cpr@holder{American Astronomical Society}
\def\received#1{\gdef\@recvdate{#1}} \received{\relax}
\def\revised#1{\gdef\@revisedate{#1}} \revised{\relax}
\def\accepted#1{\gdef\@accptdate{#1}} \accepted{\relax}
\def\journalid#1#2{\gdef\@jourvol{#1}\gdef\@jourdate{#2}}
\def\articleid#1#2{\gdef\@startpage{#1}\gdef\@finishpage{#2}}
\def\paperid#1{\gdef\@paperid{#1}} \paperid{MS-0001-SAMP}
\def\ccc#1{\gdef\CCC@code{#1}} \ccc{000-00\$75.95-CDB}
\def\cpright#1#2{\@nameuse{cpr@#1} \gdef\cpr@year{#2}
\typeout{`#1' copyright \cpr@year.}}
\newcount\@cprtype \@cprtype=\@ne
\def\cpr@AAS{\@cprtype=1}
\def\cpr@PD{\@cprtype=2}
\def\cpr@Crown{\@cprtype=3}
\def\cpr@none{\@cprtype=4}
\def\cpr@ASP{\@cprtype=5}
\def\cpr@year{\number\year}
\def\@slug{\par\noindent
\ifcase\@cprtype
	\relax
\or
	Copyright \cpr@year\space by the \cpr@holder.
\or
	This article is in the public domain.
\or
	Crown copyright \cpr@year\space by the \cpr@holder.
\or
	No copyright is claimed for this article.
\or
	Copyright \cpr@year\space by the ASP.
\fi
\par\noindent
Manuscript number \@paperid.\par\noindent
\CCC@code
}
\def\lefthead#1{\gdef\@versohead{#1}} \lefthead{\relax}
\def\righthead#1{\gdef\@rectohead{#1}} \righthead{\relax}
\def\@runheads{\@tempcnta\c@page
\@whilenum \@tempcnta >0\do{
\vskip 3ex
\hbox to30pc{\small\expandafter\uppercase\expandafter{\@versohead}:
	\expandafter\uppercase\expandafter{\@rectohead}\hfil}
\advance\@tempcnta by\m@ne}
}
\def\slugcomment#1{\gdef\slug@comment{#1}} \slugcomment{}
\newdimen\@slugcmmntwidth \@slugcmmntwidth .67\textwidth
\long\def\@makeslugcmmnt{\ifx\slug@comment\@empty\relax\else
\setbox\@tempboxa\hbox{\slug@comment}
\ifdim \wd\@tempboxa >\@slugcmmntwidth
\hbox to\textwidth{\hss
	    \parbox\@slugcmmntwidth\slug@comment}
\else
\hbox to\textwidth{\hfil\box\@tempboxa}
\fi
\vskip 2ex
\fi}
\def\@rcvaccrule{\vrule\@width1.75in\@height0.5pt\@depth\z@}
\def\@dates{{Received}\space%
\if\@recvdate\relax\@rcvaccrule\else\@recvdate\fi;%
\hspace{1.5em}{accepted}\space%
\if\@accptdate\relax\@rcvaccrule\else\@accptdate\fi%
}
\def\sluginfo{{\center
\@dates

\endcenter}}

\def\abstract{
\begin{center}
{\bf{ABSTRACT}}
\end{center}
\quotation
}
\def\title#1{\@makeslugcmmnt{\center\large\bf{#1}\endcenter}
\thispagestyle{empty}}
\def\author#1{{\topsep\z@\center\normalsize#1\endcenter}}
\let\authoraddr=\@gobble
\def\affil#1{\vspace*{-2.5ex}{\topsep\z@\center#1\endcenter}}

\def\and{\vspace*{-0.5ex}{\topsep\z@\center and\endcenter}}
\def\@keywordtext{Subject headings}
\def\@keyworddelim{---}
\def\keywords#1{\vspace*{-.7ex}
\if@twocolumn\noindent{{\it\@keywordtext:\/}\space\@kwds{#1}}
\else{\quote{\it\@keywordtext:\/}\space\@kwds{#1}\endquote}
\fi}

\def\@kwds#1{#1\relax}
\skip\footins 4ex plus 1ex minus .5ex
\newif\if@firstsection \@firstsectiontrue
\def\section{\if@firstsection
\@firstsectionfalse\fi
\@startsection {section}{1}{\z@}
{5ex plus .5ex}{1ex plus .2ex}{\normalsize\bf}}
\def\subsection{\@startsection{subsection}{2}{\z@}
{5ex plus .5ex}{1ex plus .2ex}{\normalsize\bf}}
\def\subsubsection{\@startsection{subsubsection}{3}{\z@}
{5ex plus .5ex}{1ex plus .2ex}{\normalsize\it}}
\def\thesection{\@arabic{\c@section}.}
\def\thesubsection{\thesection\@arabic{\c@subsection}.}
\def\thesubsubsection{\thesubsection\@arabic{\c@subsubsection}.}
\def\theparagraph{\thesubsubsection\@arabic{\c@paragraph}:}

\let\acknowledgements=\acknowledgments
\def\@sect#1#2#3#4#5#6[#7]#8{\ifnum #2>\c@secnumdepth
\def\@svsec{}\else
\refstepcounter{#1}\edef\@svsec{\csname the#1\endcsname\hskip 1em }\fi
\@tempskipa #5\relax
\ifdim \@tempskipa>\z@
\begingroup \center#6\relax
\@hangfrom{\hskip #3\relax\@svsec}{\interlinepenalty \@M
	  \sec@upcase{#8}\par}%
\endcenter\endgroup
\csname #1mark\endcsname{#7}\addcontentsline
{toc}{#1}{\ifnum #2>\c@secnumdepth \else
\protect\numberline{\csname the#1\endcsname}\fi
#7}\else
\def\@svsechd{#6\hskip #3\@svsec \sec@upcase{#8}\csname #1mark\endcsname
{#7}\addcontentsline
{toc}{#1}{\ifnum #2>\c@secnumdepth \else
\protect\numberline{\csname the#1\endcsname}\fi
#7}}\fi
\@xsect{#5}}
\def\@ssect#1#2#3#4#5{\@tempskipa #3\relax
\ifdim \@tempskipa>\z@
\begingroup #4\center\@hangfrom{\hskip #1}{\interlinepenalty \@M
\sec@upcase{#5}\par}\endcenter\endgroup
\else \def\@svsechd{#4\hskip #1\relax \sec@upcase{#5}}\fi
\@xsect{#3}}
\def\appendix{\par
\setcounter{section}{0}
\setcounter{subsection}{0}
\setcounter{equation}{0}
\def\thesection{\Alph{section}.}
\def\theequation{\hbox{\normalsize\Alph{section}\arabic{equation}}}}
\newcounter{cureqno}
{\let\theequation\@curtheeqn%
\setcounter{equation}{\value{cureqno}}}
\def\eqnum#1{\def\theequation{#1}\let\@currentlabel\theequation
\addtocounter{equation}{\m@ne}}
\def\references{\subsection*{REFERENCES}
\bgroup\parindent=\z@\parskip=0pt
\def\refpar{\par\hangindent=3em\hangafter=1}}
\def\endreferences{\refpar\egroup\revtex@pageid}
\def\thebibliography{\subsection*{REFERENCES}
\list{\null}{\leftmargin 3em\labelwidth\z@\labelsep\z@\itemindent -3em
\usecounter{enumi}}
\def\refpar{\relax}
\def\newblock{\hskip .11em plus .33em minus .07em}
\sloppy\clubpenalty4000\widowpenalty4000
\sfcode`\.=1000\relax}
\def\endthebibliography{\endlist\revtex@pageid}
\def\@biblabel#1{\relax}
\def\@cite#1#2{#1\if@tempswa , #2\fi}
\def\reference{\relax\refpar}

\def\@citex[#1]#2{\if@filesw\immediate\write\@auxout{\string\citation{#2}}\fi
\def\@citea{}\@cite{\@for\@citeb:=#2\do
{\@citea\def\@citea{,\penalty\@m\ }\@ifundefined
{b@\@citeb}{\@warning
{Citation `\@citeb' on page \thepage \space undefined}}%
{\csname b@\@citeb\endcsname}}}{#1}}
\def\tablenotemark#1{\rlap{$^{\rm #1}$}}
\newtoks\@temptokenb
\def\tblnote@list{}
\def\tablenotetext#1#2{
\@temptokena={\vspace{.5ex}{\noindent\llap{$^{#1}$}#2}\par}
\@temptokenb=\expandafter{\tblnote@list}
\xdef\tblnote@list{\the\@temptokenb\the\@temptokena}}
\def\spew@tblnotes{
\ifx\tblnote@list\@empty\relax
\else
\vspace{4.5ex}
\footnoterule
\vspace{.5ex}
{\footnotesize\tblnote@list}
\gdef\tblnote@list{}
\fi}
\def\endtable{\spew@tblnotes\end@float}
\@namedef{endtable*}{\spew@tblnotes\end@dblfloat}
\let\tableline=\hline
\long\def\@makecaption#1#2{\vskip 2ex\noindent #1 #2\par}
\def\tablenum#1{\def\thetable{#1}\let\@currentlabel\thetable
\addtocounter{table}{\m@ne}}
\def\figurenum#1{\def\thefigure{#1}\let\@currentlabel\thefigure
\addtocounter{figure}{\m@ne}}
\newbox\pt@box
\newdimen\pt@width
\newcount\pt@line
\newcount\pt@nlines
\newcount\pt@ncol
\def\colhead#1{\omit\hidewidth{#1}\hidewidth\global\advance\pt@ncol by\@ne}
\def\tablecaption#1{\gdef\pt@caption{#1}} \def\pt@caption{\relax}
\def\tablehead#1{\gdef\pt@head{\hline\hline\relax\\[-1.7ex]
#1\hskip\tabcolsep\\[.7ex]\hline\relax\\[-1.5ex]}} \def\pt@head{\relax}
\def\tabletail#1{\gdef\pt@tail{#1}} \def\pt@tail{\relax}
\def\tablewidth#1{\pt@width=#1} \pt@width\textwidth
\def\tableheadfrac#1{\gdef\pt@headfrac{#1}} \def\pt@headfrac{.1}
\def\pt@calcnlines{\@tempdima\pt@headfrac\textheight
\@tempdimb\textheight\advance\@tempdimb by-\@tempdima
\@tempdima\arraystretch\baselineskip
\divide\@tempdimb by\@tempdima
\global\pt@nlines\@tempdimb}
\def\pt@tabular{\hbox \bgroup $\let\@acol\@ptabacol
\let\@classz\@tabclassz
\let\@classiv\@tabclassiv \let\\\@tabularcr\@tabarray}
\def\@ptabacol{\edef\@preamble{\@preamble \hskip \tabcolsep\tabskip\fill}}
\def\fnum@ptable{Table \thetable}
\def\fnum@ptablecont{Table \thetable---{\rm Continued}}
\def\set@mkcaption{\long\def\@makecaption##1##2{
\center\rm##1.\quad##2\endcenter\vskip 2.5ex}}
\def\set@mkcaptioncont{\long\def\@makecaption##1##2{
\center\rm##1\endcenter\vskip 2.5ex}}
{\crcr\noalign{\vskip .7ex}\hline\endtabular%
\pt@width\wd\pt@box\box\pt@box\spew@ptblnotes%
\typeout{Table \thetable\space has been set to width \the\pt@width}%
\endcenter\end@float}
\def\startdata{\pt@line=0\pt@calcnlines%
\ifdim\pt@width>\z@\def\@halignto{to \pt@width}\else\def\@halignto{}\fi%
\let\fnum@table=\fnum@ptable\set@mkcaption%
\@float{table}\center\caption{\pt@caption}\leavevmode%
\setbox\pt@box=\pt@tabular{\pt@format}\pt@head}
\def\pt@nl{\global\advance\pt@line by\@ne%
\ifnum\pt@line=\pt@nlines%
\endtabular\box\pt@box
\endcenter\end@float\clearpage%
\addtocounter{table}{\m@ne}%
\let\fnum@table=\fnum@ptablecont\set@mkcaptioncont%
\@float{table}\center\caption{\pt@caption}\leavevmode%
\global\pt@ncol=0%
\setbox\pt@box=\pt@tabular{\pt@format}\pt@head%
\global\pt@line=0%
\else\\
\fi}
\let\nl=\pt@nl
\let\nextline=\pt@nl

\def\tablebreak{\pt@line\pt@nlines\advance\pt@line by\m@ne\pt@nl}
\def\cutinhead#1{\noalign{\vskip 1.5ex}
\hline\pt@nl\noalign{\vskip -2.0ex}
\multicolumn{\pt@ncol}{c}{#1}\pt@nl
\noalign{\vskip .8ex}
\hline\pt@nl\noalign{\vskip -2ex}}
\def\sidehead#1{\noalign{\vskip 1.5ex}
\multicolumn{\pt@ncol}{@{\hskip\z@}l}{#1}\pt@nl
\noalign{\vskip .5ex}}
\def\set@tblnotetext{\def\tablenotetext##1##2{{%
\@temptokena={\vspace{0ex}{%
\parbox{\pt@width}{\hskip1em$^{\rm ##1}$##2}\par}}%
\@temptokenb=\expandafter{\tblnote@list}
\xdef\tblnote@list{\the\@temptokenb\the\@temptokena}}}}
\def\spew@ptblnotes{
\ifx\tblnote@list\@empty\relax
\else
\par
\vspace{2ex}
{\tblnote@list}
\gdef\tblnote@list{}
\fi}
\def\tablerefs#1{\@temptokena={\vspace*{3ex}{%
\parbox{\pt@width}{\hskip1em\rm References. --- #1}\par}}%
\@temptokenb=\expandafter{\tblnote@list}
\xdef\tblnote@list{\the\@temptokenb\the\@temptokena}}
\def\tablecomments#1{\@temptokena={\vspace*{3ex}{%
\parbox{\pt@width}{\hskip1em\rm Note. --- #1}\par}}%
\@temptokenb=\expandafter{\tblnote@list}
\xdef\tblnote@list{\the\@temptokenb\the\@temptokena}}
\def\thefigure{\@arabic\c@figure}
\def\fnum@figure{{\rm Fig.\space\thefigure.---}}
\def\thetable{\@arabic\c@table}
\def\fnum@table{{\rm Table \thetable:}}
\def\fps@figure{bp}
\def\fps@table{bp}
\@ifundefined{epsfbox}{\@input{epsf.sty}}{\relax}
\def\plotone#1{\centering \leavevmode
\epsfxsize=\columnwidth \epsfbox{#1}}

\let\jnl@style=\rm
\def\ref@jnl#1{{\jnl@style#1}}
\def\aj{\ref@jnl{AJ}}
\def\araa{\ref@jnl{ARA\&A}}
\def\apj{\ref@jnl{ApJ}}
\def\apjl{\ref@jnl{ApJ}}
\def\apjs{\ref@jnl{ApJS}}
\def\ao{\ref@jnl{Appl.Optics}}
\def\apss{\ref@jnl{Ap\&SS}}
\def\aap{\ref@jnl{A\&A}}
\def\aapr{\ref@jnl{A\&A~Rev.}}
\def\aaps{\ref@jnl{A\&AS}}
\def\azh{\ref@jnl{AZh}}
\def\baas{\ref@jnl{BAAS}}
\def\jrasc{\ref@jnl{JRASC}}
\def\memras{\ref@jnl{MmRAS}}
\def\mnras{\ref@jnl{MNRAS}}
\def\pra{\ref@jnl{Phys.Rev.A}}
\def\prb{\ref@jnl{Phys.Rev.B}}
\def\prc{\ref@jnl{Phys.Rev.C}}
\def\prd{\ref@jnl{Phys.Rev.D}}
\def\prl{\ref@jnl{Phys.Rev.Lett}}
\def\pasp{\ref@jnl{PASP}}
\def\pasj{\ref@jnl{PASJ}}
\def\qjras{\ref@jnl{QJRAS}}
\def\skytel{\ref@jnl{S\&T}}
\def\solphys{\ref@jnl{Solar~Phys.}}
\def\sovast{\ref@jnl{Soviet~Ast.}}
\def\ssr{\ref@jnl{Space~Sci.Rev.}}
\def\zap{\ref@jnl{ZAp}}

\def\lesssim{\mathrel{\hbox{\rlap{\hbox{\lower4pt\hbox{$\sim$}}}\hbox{$<$}}}}
\def\gtrsim{\mathrel{\hbox{\rlap{\hbox{\lower4pt\hbox{$\sim$}}}\hbox{$>$}}}}

\def\ion#1#2{#1$\;${\small\rm\@Roman{#2}}\relax}

\newcount\lecurrentfam
\def\LaTeX{\lecurrentfam=\the\fam \leavevmode L\raise.42ex
\hbox{$\fam\lecurrentfam\scriptstyle\kern-.3em A$}\kern-.15em\TeX}
\def\sizrpt{
(\fontname\the\font): em=\the\fontdimen6\font, ex=\the\fontdimen5\font
\typeout{
(\fontname\the\font): em=\the\fontdimen6\font, ex=\the\fontdimen5\font
}}
\makeatother
\tightenlines
\newcommand{\etal}{{\em et al.~}}
\newcommand{\kms}{$km\;s^{-1}\;$}
\newcommand{\ppp}{^\prime}
\def\lsim{\lower 2pt \hbox{$\, \buildrel {\scriptstyle <}\over
{\scriptstyle \sim}\,$}}
\def\gsim{\lower 2pt \hbox{$\, \buildrel {\scriptstyle >}\over
{\scriptstyle \sim}\,$}}
\begin{document}
\title{Exponential Tails in the Centroid Velocity Distributions of
Star-Forming Regions}
%
\author{Mark S. Miesch}
\affil{Joint Institute for Laboratory Astrophysics \\
and Department of Astrophysical, Planetary, and Atmospheric Sciences \\
University of Colorado -- Campus Box 440
Boulder, CO 80309 \\ miesch@solarz.colorado.edu}

\and

\author{John M. Scalo}
\affil{Department of Astronomy \\
University of Texas at Austin
Austin, TX 78712 \\ parrot@astro.as.utexas.edu}

\vspace{0.5in}
\affil{(a shorter version) Submitted to the Astrophysical Journal Letters}
\affil{Draft date: Dec. 1, 1994}

\keywords{ISM: clouds, ISM: molecules, stars: formation, turbulence}
\lefthead{Miesch and Scalo}
\righthead{Centroid Velocity Distributions of SF regions}
\begin{abstract}
Probability density functions (pdfs) of $^{13}CO$ emission line centroid
(line-of-sight, intensity-weighted average) velocities are presented
for several densely sampled molecular clouds as quantitative descriptors
of their underlying dynamics.  Although some are approximately Gaussian
in form, most of the pdfs exhibit relatively broader, often nearly
exponential, tails, similar to the pdfs of velocity {\em
differences} and {\em derivatives} (but not the velocity field itself)
found in experiments and numerical simulations of incompressible
turbulence.  The broad pdf tails found in the present
work are also similar to those found in decades-old measurements
of interstellar velocity pdfs using atomic line centroids,
and to the excess wing emission
recently found in individual molecular line profiles.
Some possible interpretations of the observed deviations are
briefly discussed, although none of these account for the nearly
exponential tails.
\end{abstract}
\normalsize
\renewcommand{\theenumi}{\roman{enumi}}

\section{Introduction}

         Although a great deal of effort has been devoted to quantitatively
describing the complex spatial structure of star-forming regions (for
recent approaches see Falgarone and Phillips 1990, 1991, Gill and Henriksen
1991,
Langer, Wilson and Anderson 1993, Wiseman and Adams 1994a,b, Zimmermann and
Stutzki 1994, Houlahan and Scalo 1992, Scalo 1990, Chappell and Scalo
1994), comparatively
little attention has been paid to characterizing the radial velocity
dimension of the data except for
studies of possible velocity dispersion-size scaling relations (see
Falgarone, Puget, and Perault 1992 and references given there),
 estimation of the velocity correlation
function and related 2-point statistics (see Hobson 1992, Kitamura et al.
1993, Miesch and Bally 1994 and references to earlier work given
there) and searches for evidence of rotation (e.g. Goodman et al. 1994).
Since one expects a signature of the dynamical and physical processes
to appear in the velocity field,  it is surprising and unfortunate that
studies of this kind are not more prevalent.
As a step toward a better
understanding of molecular cloud velocity structure,
Falgarone and coworkers (Falgarone
1989, Falgarone and Phillips 1990,1991, Falgarone et al. 1994; see below
for discussion) have explicitly tried to relate radial velocity
information to dynamical processes through the comparison of
observed line
profiles with frequency distributions, or probability distribution
functions, found in experimental and simulation studies of turbulence.  The
present paper tries to extend that program to the frequency distribution of
centroid velocities.

 Recent work in several areas suggests that the one-point
probability distribution function (pdf, or, loosely, the histogram)  of
dynamical variables like velocity is a useful tool that may be sensitive to
dynamical processes.  These studies include large scale structure of galaxy
velocities (Bernardeau 1994,  Korman et al. 1994, Catelan and Moscardini 1994),
incompressible terrestrial turbulence (see below),  distinguishing
nonlinear chaotic processes from stochastic processes (Wright and Schult
1993), and characterization of samples of musical volume fluctuations
(Scalo and Chappell 1995).  In particular, studies of incompressible
turbulence have shown that the higher moments (skewness, kurtosis,...) of
the pdf can be used to constrain physical models for turbulent
intermittency.  For incompressible turbulence the pdf of the velocity field
itself is very nearly
Gaussian, at least on large enough scales (Batchelor 1953 and Monin and Yaglom
1971 review early work; see more recent experiments and simulations in
Anselmet, Gagne, and Hopfinger 1984, Figure 1; Kida and Murakami 1989, Figure
6;
Jayesh and Warhaft 1991, Figure 1; Chen et al. 1993, Figure 3.), although
non-zero skewness must exist at some level in order to provide
energy transfer among different scales.
Non-Gaussian behavior is well-established for velocity
{\em differences} at small scales and velocity
{\em derivatives}, and there is strong evidence from experiments
and simulations for non-Gaussian behavior in many other variables (see the
papers referred to above and Chen et al. 1989, Castaing, Gagne and Hopfinger
1990,  Vincent and Meneguzzi 1991, She et al. 1993).  Often the pdf of the
velocity difference or derivative field exhibits a near-exponential
behavior at smaller and smaller scales, and much work has gone into
understanding this behavior physically, especially in terms of the
stretching properties of the advection operator (see She 1991 for a
review).   Part of the motivation of the present work is to
investigate whether any of these properties occur in the more complex
``turbulence'' of interstellar clouds, and whether even the velocity
fluctuation field itself presents measurable deviations from a Gaussian
pdf.

 Falgarone and Phillips (1990, 1991) have shown that line profiles
constructed from high-sensitivity CO molecular line data (in several
transitions) exhibit excess wing emission, relative to a single Gaussian,
over a very large range of scales, from 0.02 to 450 pc .  For all these
line profiles the width of the wings is about 3 times the width of the line
core if both are fit by Gaussians, but the fractional intensity of the wing
component (fraction of mass at high velocities) varies between about 0.03
and 0.8.   Broad wings were also found in high latitude molecular clouds by
Magnani, Blitz and Wendel (1989).  The presence of similar broad wings in
regions
whose scales are gravitating and non-self-gravitating, and in regions with and
without internal massive star formation,
suggests that the behavior is not due to stellar winds or collapse motions, and
the variation
in wing width  in these regions seems to rule out a dilute warm gaseous
component, as pointed out by Falgarone and Phillips.
Since the line profile, in the optically thin case, is in effect
a histogram of radial velocities, the broad wings have been viewed
in the context of non-Gaussian pdfs, although there is some confusion
concerning whether the line profile should be interpreted as the pdf of
average line of sight
velocities or of velocity differences; the latter interpretation is adopted
by Falgarone and Phillips (1990) in comparisons with laboratory data.

 A number of physically-based explanations for the broad wings have
been proposed, as discussed by Falgarone et al. 1994 and in $\S4$ below,
but the systematics of non-Gaussian
pdfs in star-forming regions have yet to be established.  It is not even
clear that line profiles give a valid representation of the pdf.  Every
line profile samples a line-of-sight velocity field which in general
contains a component whose characteristic scale is a significant fraction
of the sample depth.  The
form of these systematic line-of-sight motions is unknown and may severely
limit the correspondence between the line profile and velocity pdf.
Such problems can largely be circumvented in analyses of simulations,
where it is possible to to insure homogeneity on the largest scales
(as in Falgarone et al. 1994), but homogeneity is probably not a
good assumption in general for interstellar clouds.
It is not difficult to show that the addition of a systematic component
can significantly alter the estimate of the distribution of the velocity
{\em fluctuations}, which is the function of interest.  A cloud in
non-solid-body
rotation about its center, for example, will yield non-Gaussian line
profiles along lines of sight displaced from the projection of the rotation
axis onto the plane of the sky (provided this projection is nonzero).  In
particular, these profiles will exhibit apparent excess wing emission due
solely to the {\em smearing} arising from the variation of the
line-of-sight component of the rotational velocity with depth in the cloud,
which will thus distort the pdf of velocity fluctuations.
In addition, radiative transfer effects can distort emission lines and
cause the wings to become relatively more prominent if the cloud is
optically thick (although Falagarone \& Phillips 1990 argue against
this interpretation of the broad wings on the basis of their observed
shapes).

 An alternative procedure, which we adopt here, is to estimate the
pdf of {\em centroid line velocities}  (intensity weighted average velocity
along the line of sight) sampled over a densely observed individual star
formation region.
While a  ``line profile'' is a measure of the radial velocity (or
velocity difference) pdf sampled along the line of sight,
either in a single beam or averaged
over many beams, the ``centroid pdf'' is the pdf of the mean velocity of
line profiles taken over a large spatial sample of positions in the
plane of the sky.  The two functions differ in the direction along which the
sampling for the pdf is taken, and in the quantity sampled.
It follows from the central limit theorem that any deviations from
Gaussian behavior in the centroid pdf implies the
existence of higher-point spatial correlations in the velocity field,
an interpretation which also applies to the line profile if it is viewed
as an average over the beam.
The advantages of the centroid pdf approach include the much lower
sensitivity required for each of the individual line profiles and the
weaker dependence of the results on large scale, systematic motions which,
although still a concern, will tend to be mitigated by the line-of-sight
averaging and by space filtering of the velocity maps (see below).  For
example, the centroid velocities of the rotating cloud discussed above will
vary in an obvious way with position, and the effects of rotation can
therefore be removed by applying an appropriate filter, a procedure which
is not possible with the individual profiles unless the rotation curve of
the cloud is known. In addition, the presence of a warm ``interclump'' medium,
or of ``optical depth broadening'', which would both contribute
to the line profiles, will not much affect
the pdf of centroid velocities, since the thermal component and the
line saturation are
symmetric (although the centroid pdf, in the optically thick case,
would only sample fluctuations on the leading edge of the cloud).
The problem with this approach is that the number of
velocities (positions) which must be sampled in order to
accurately estimate the tails of the pdf is very large, at least of order
1000.  Furthermore, the relationship between the pdf of an average
line-of-sight quantity (centroid velocity in this case) and the pdf of the
radial velocity distribution in three dimensions is not clear; in the case
of line profiles the third dimension is sampled, but the systematic motions
in this dimension cannot be removed.

 Work aimed at determining the pdf of interstellar gas motions based on
centroid velocities dates
back to the early 1950s.  Several studies used optical absorption line
velocities of ``clouds'' along the line of sight to relatively nearby OB
stars (Blaauw 1952, Huang 1950, Kaplan 1954, Takakubo 1958, Munch 1957) and
velocities of HI 21cm emission (Takakubo 1967, Mast and Goldstein 1970) and
absorption (Crovisier 1978) lines to estimate the residual, or peculiar,
velocity distribution, after correction for solar motion and differential
galactic rotation.  These studies all refer to fairly local gas, with
distances less than about 500-1000 pc.  In addition, Verschuur (1974)
presented detailed HI emission mapping and ``cloud'' identification (from the
spatial-velocity data cube) in two regions, one of which is suitable for
estimation of the velocity pdf.   In reading these studies, one finds either
what the authors consider as clear evidence for non-Gaussian velocity
distributions (Huang 1950, Blaauw 1952, Kaplan 1954, Munch 1957, Takakubo
1958, Mast and Goldstein 1970), with an exponential function giving a
better fit to the pdfs, or a concern that systematic effects and small
number statistics prevent a conclusion concerning the functional form of
these distributions (Takakubo 1967, Crovisier 1978, Verschuur 1974).  In no
case does one find evidence supporting a Gaussian velocity distribution.
However  no theoretical explanation of these surprising results was
forthcoming, and with the advent and subsequent prominence of molecular
line observations, these studies were never repeated and were in effect
forgotten.  We know of no previous molecular line studies that have
examined the centroid
velocity pdf, that is, the histogram of centroid velocities
sampled along many lines of sight.

The present Letter examines the frequency distribution of over
29,000 independent centroid radial velocities,
divided between 12 star-forming regions
and subregions, derived from the Bell Labs $^{13}CO$ (1-0) survey.  We briefly
discuss the data, present the pdfs estimated as histograms, estimate the
skewness and kurtosis for each region, and briefly discuss the theoretical
relevance of the results, as well as give a brief summary of
theoretical models.
A more complete presentation, including a discussion of the pdfs of velocity
differences and
linewidths, will be given elsewhere (Miesch and Scalo 1995, hereafter paper
II).

\section{Data}

The data sets used here are the same as those used by Miesch \&
Bally (hereafter MB) in their study of the fluctuating, or ``turbulent'',
velocity structure in selected nearby molecular clouds as characterized by
two-point statistical functions.  The observations are of $^{13}$CO J=1-0
emission in L1228, L1551, Mon R2, and Ori B as well as $^{13}$CO J=2-1
observations of the molecular
gas surrounding the Herbig-Haro object HH83 which lies in the Orion A cloud,
just west of L1641 (Bally \etal 1994).   We thank John Bally for making these
data
available  to us for the present work.
Notice that all of these regions  and subregions are actively
forming stars, as evidenced by observed outflows, HH objects, reflection
nebulae,
maser emission, probable T-Tauri and related stars, and high IR fluxes
and point source counts as revealed by IRAS, and that energy input
from these young stars has likely had a significant dynamical influence on
the density and velocity structure of the clouds (Bally \& Devine 1994;
Bally, Castets, \& Duvert 1994; Xie 1992; Pound \& Bally 1991;
Reipurth \& Olberg 1991; Ogura \& Sato 1990; Bally \etal 1989;
Genzel \& Stutzki 1989; Haikala \& Laurejs 1989; see paper II for
further discussion).  In addition, external energy sources such as
the OB associations in Orion (Genzel \& Stutzki 1989) and the
probable supernova remnant associated with L1228 (Grenier \etal 1989)
could supply a significant amount of mechanical energy, both directly
through transmitted shock waves (Miesch \& Zweibel 1994)
and ablation flows, and by
induced star formation.  The highly
``stirred-up'' nature of these regions is in contrast to some of the other
regions in which broad emission line wings have been found
(a quantitative measure of the
relative star-forming activity of each region will be given in paper II).

        All but the HH83 observations were made with the AT\&T
Bell Laboratories 7-m Crawford hill antenna (with a beamwidth of
1$\ppp$.7 at 110 GHz) as part of the Bell Labs molecular cloud survey,
while the HH83 data were obtained with the IRAM 30-m telescope on Pico
Veleta, Spain (with a 0$\ppp$.22 beamwidth).  The velocity resolution was
0.26 \kms for L1228, Mon R2, and Ori B, and 0.13 \kms for L1551 and HH83.
Two regions in particular, Mon R2 and Ori B, were clearly composed of
several distinct clouds, or subregions, (3 and 6 respectively) which were
separated out in space and velocity prior to analysis (regions
1a, 1b, and 1c in Ori B are small clouds to the northeast of L1627,
which lies in region 2.  Region 4 is southwest of that,
near the Horsehead Nebula and
includes L1630, and region 3 spatially overlaps regions 2 and 4, but is
at a significantly lower velocity).

The velocity
centroid (defined as $\sum u_i T_i/\sum T_i$, where $T_i$ and $u_i$ are the
brightness temperature and the velocity of the $i^{th}$ channel and
summation is over the line profile) was computed along each line of sight
for which both the integrated intensity and the peak brightness temperature
exceeded imposed threshold values, intended to minimize the influence of
noise-dominated spectra.  However it should be remembered in what follows
that the pdfs are biased by what is probably a column density cutoff, as of
course are all published descriptions of cloud structure.
         The data were oversampled by a factor of two when constructing the
velocity centroid maps, so the actual number of independent spectra used
for each of these regions is about a factor of four smaller than the total
number of points in the map.  This latter number (the number of points) is
listed in Table 1 for each region we have studied, and varies from 822 to
21876.  For further
information on how the centroid velocity maps were constructed, and for a
more detailed description of the observations themselves, see MB and the
references given at the beginning of this section.
Images of the velocity centroid
structure will be presented in paper II.

\hoffset4in
\small
\begin{table}[tb]
\centering
\begin{center}
{\bf Table 1} \\ PDF Attributes \protect\tablenotemark{a}
\end{center}
\begin{tabular}{lllllll}
\tableline\tableline
Region & No. of & Filtering & Core & Dispersion & Skewness & Kurtosis \\
 & points\tablenotemark{b} & Ratio & Fraction \\
 & $N$ & $\sqrt{N}/\ell_f$ & (percent) & $\sigma$ ($km\;s^{-1}$) & $\xi$
& $\kappa$ \\\tableline
HH83 & 6235 & 2.4 & 91 (85) & 0.12 (0.12) & -0.62 (-0.53) & 4.2 (3.9) \\
L1228\tablenotemark{c} & 3532 & 3.5 & 96 (94,95) & 0.14 (0.15, 0.14) & -0.20
(-0.33, -0.31) & 5.9 (12, 7.4)\\
L1551 & 4726 & 2.7 & 79 (71) & 0.18 (0.22) & 0.65 (-0.54) & 4.6 (7.0) \\
Mon R2 \hspace{.05in}1 & 8630 & 2.3 & 100 (100) & 0.37 (0.38) & -0.14 (0.00) &
3.0
(3.0) \\
\hspace{.605in} 2 & 7592 & 2.1 & 100 (100) &  0.23 (0.23) & -0.37 (0.00) & 3.2
(3.0) \\
\hspace{.605in} 3 & 12208 & 2.7 & 100 (100) & 0.29 (0.29) & 0.054 (0.00) & 3.2
(3.0) \\
Ori B \hspace{.2in} 1a & 822 & 1.4 & 100 (100) & 0.15 (0.18) & 0.13 (0.00) &
3.0 (6.0) \\
\hspace{.605in} 1b & 2094 & 2.2 & 97 (100) & 0.22 (0.20) & -0.60 (0.00) & 4.8
(3.0) \\
\hspace{.605in} 1c & 4801 & 3.3 & 89 (97) & 0.37 (0.33) & -0.86 (-0.46) & 4.2
(5.7) \\
\hspace{.605in} 2\tablenotemark{d} & 18420 & 3.3 & 97 (96,96) & 0.35 (0.35,
0.31) & -1.5

(-1.4, -0.28) & 11 (12, 5.2) \\
\hspace{.605in} 3 & 10433 & 2.0 & 90 (83) & 0.64 (0.69) & 0.54\tablenotemark{e}
(0.73) & 3.9
(5.6) \\
\hspace{.605in} 4 & 21876 & 2.9 & 100 (100) & 0.47 (0.45) & 0.19 (0.00) & 4.8
(6.0) \\ \tableline
\end{tabular}
\tablenotetext{a}{The values without and within parentheses are obtained
respectively by discrete summations using the observed histograms,
and by integration using the piecewise smooth analytic fits described in the
text and shown in Fig. 1}
\tablenotetext{b}{The actual number of spectra used is
roughly one fourth these values because the maps were oversampled by
a factor of two (see text).}
\tablenotetext{c}{The two analytic values are for exponential and Gaussian
wings respectively (see Fig. 1b).}
\tablenotetext{d}{The latter analytic value excludes the
component centered at -2.25 $km\;s^{-1}$.}
\tablenotetext{e}{The skewness calculated directly from the data
points excludes all points for which $|v|\gsim 2.2$\kms (see text).}
\end{table}

\normalsize

\hoffset0in
        We are interested here in single-point statistics describing the
fluctuating velocity structure in star-forming regions, and we have therefore
filtered out large scale, systematic motions such as velocity gradients
across the maps by first convolving the raw data with a suitable smoothing
function (either a Zurflueh or moving average filter; see MB for a
detailed discussion of the filtering process) and then subtracting the
smoothed map from the original to obtain the fluctuating
velocity component.  The width of each smoothing
function (which is roughly equal to the cutoff wavelength of the high-pass
filtering process) was chosen to be the largest value at which asymmetric
lobes (which are the signature of large-scale gradients) were no longer
discernible in the autocorrelation function of the residual centroid map.
The ratio of the effective scale of the observations (taken to be
the square root of the total number of points) to the filter size is given
in Table 1 for each region.  The unfiltered probability density functions were
usually significantly different
than the filter-subtracted versions, the former showing multimodal behavior and
large
asymmetries.  This illustrates the danger of estimating pdfs (or any
statistical
descriptor) from data that contain structure whose scale is a significant
fraction of the mapped region, and also suggests, as mentioned earlier,
that spectral line profiles may not give a valid representation of the
velocity fluctuation pdf, since there is no way to filter out the
line-of-sight systematic component.  Indeed, the average line profiles
of these regions exhibit a variety of forms that suggest systematic large
scale structure along the line of sight (Miesch and Scalo 1995).

\section{Analysis and Results}

\begin{figure}[p]
\plotone{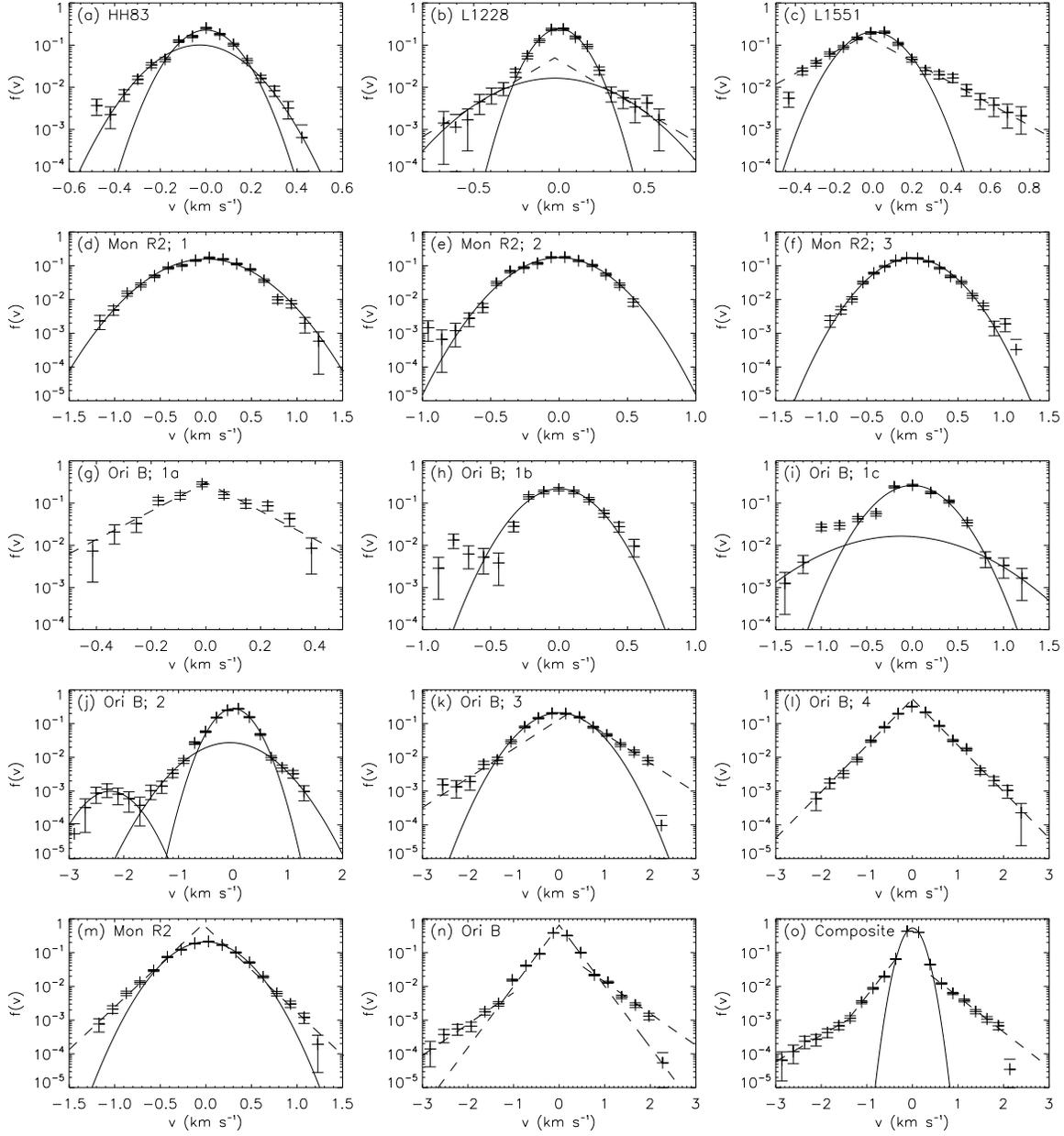}
\caption{
Probability distribution functions of the observed
centroid velocities, including both histograms (crosses, with
statistical error bars) and functional
fits (lines), the latter of which take two forms:
Gaussian
(\protect$f(v)=f_*e^{-(v-v_*)^2/2\sigma_*^2}\protect$; solid lines) and
exponential
(\protect$f(v)=f_*e^{-\protect\sqrt{2}|v-v_*|/\sigma_*}\protect$;
dashed lines).  Each region
is labeled and the final three frames represent the normalized pdfs
averaged over the three Mon R2 regions (m), the six Ori B regions (n),
and all 12 data sets (o).  Note that, for the linear-logarithmic
axes chosen, an exponential takes the form of a straight line and
a Gaussian a parabola.  The anomalously low point in the high velocity
tail of Ori B, region 3 (k), which is reflected in the composites (n)
and (o), is probably an artifact, as explained in the text.
}
\end{figure}

        Histograms (hereafter probability density functions or pdfs) of the
centroid velocities for each of the 12 regions we have studied are
exhibited in Figure 1, normalized with respect to the total number of
points in each map.
The plots are log-linear, so a Gaussian will appear
as a parabola while an exponential will appear linear.
Also included are composite histograms made by averaging the normalized
pdfs for the three Mon R2 regions (Figure 2m), for the six Ori B regions
(Figure 2n), and for all 12 data sets (Figure 2o), intended in some sense to
approximate ensemble averages (provided similar physical processes are
occurring in each region).  Observed values are shown as crosses, with
vertical error bars representing the statistical uncertainty in each bin,
which we have taken to be $\epsilon_i = 2\sqrt{N_i}/N$, where $N_i$ is the
number of events in the bin and $N=\sum_i N_i$ is the total number of
points in the map.  The denominator in this expression for $\epsilon_i$
arises from the normalization chosen and the factor of 2 is present
because, as mentioned above, these maps have been oversampled, so the
actual number of independent spectra is about one fourth the number of
points, both in the map as a whole and in the individual bins.
Overlayed on the data of Figure 1 are several functional fits
to each of the pdfs, having either Gaussian (solid lines)
or exponential (dashed lines) forms
(more general forms will be considered in paper II).
The horizontal
axes in each frame of Figure 1 have been adjusted so that $v=0$ corresponds
to the center of the fitted primary (core) component.
Table 1 includes the
fraction of the total area under each pdf curve due only to the
core component, expressed as a percentage.
 Results for Ori B, region 2 are listed both with and without the
 secondary component centered at
 at a relative velocity of -2.25\kms because this
component is likely due to residual emission from region
3, which lies in the same area of the sky but at a lower velocity.

        A useful way to quantify the shape of probability density functions
 which has been used extensively in the study of intermittency in
 incompressible turbulence (e.g. Monin \& Yaglom 1971; Vincent \&
 Meneguzzi 1991) is by means of $n^{th}$ order moments which,
 for a pdf $f(v)$, are given by
 \begin{equation}
 {\cal M}^{(n)} \equiv
 \frac{\int_{-\infty}^{\infty}v^nf(v)dv}{\int_{-\infty}^\infty f(v)dv}.
 \end{equation}
 We then define the dispersion, or standard deviation, $\sigma$, the
 skewness, $\xi$, and the kurtosis, or flatness factor, $\kappa$,
 in terms of the second, third, and fourth moments:
 \begin{equation}
 \sigma \equiv \sqrt{{\cal M}^{(2)}}, \mbox{\hspace{.3in}}
 \xi \equiv \frac{{\cal M}^{(3)}}{\sigma^3}, \mbox{\hspace{.1in} and
 \hspace{.1in}} \kappa \equiv \frac{{\cal M}^{(4)}}{\sigma^4}.
 \end{equation}
 The units of $\sigma$ are those of velocity, while $\xi$ and
 $\kappa$ are dimensionless.  Table 1 contains the values of
 $\sigma$, $\xi$, and $\kappa$ for each region, including both
 results obtained from the observed histograms using a discrete
 summation analogous to equation (1) and those obtained by analytically
 integrating the functional fits using the forms and parameters shown
 in Figure 1.  For those pdfs which appear to be composed of more than one
component,
 we have adopted a piece-wise smooth representation which changes character
 discontinuously where the tail and core fits intersect.
 When the fits are adjusted to avoid discontinuities in the first
derivative, the resulting changes in the moments of Table 1 are minimal
(if the smoothing scale at each discontinuity is the size of a bin,
these changes do not exceed 4\%).

The composite regions have been omitted from Table 1
because some quantitative results (e.g. excessively large values of the
kurtosis) arise simply as a result of averaging (or
equivalently, summing) a small number of
normalized pdfs with different dispersions, and
they are therefore not useful diagnostics of the flow fields
(although the integral over a large number of Gaussian pdfs with
fluctuating dispersions has been used by Castaing et al. 1990
to study velocity difference pdfs).  However,
we do not see how this averaging could give rise to the exponential
tails so prevalent in all of the composite pdfs, so we must regard
their general functional forms to be a result of physical processes, and as
such, to be of interest. The skewness of the composites is also of interest
and, when calculated using the discrete
data points, is found to be -0.17 for Mon R2, 0.22 for
Ori B, and 0.16 for the ``all regions'' composite (note that, in
calculating the moments for the latter two, the anomalously low point
at $v\sim2.2$\kms was excluded from the summations, as well as
all points with a velocity less than $v\lsim -2.2$\kms, in order to
minimize the influence of the sharp cutoff at high velocities, which is
probably spurious; see below).
The corresponding values using the analytic fits are -0.19, 0.66, and -0.11
(the large discrepancy for the latter two can be attributed to the
spurious cutoff, which greatly influences the values obtained directly
from the data but does not appear in the fits).

          Before proceeding, we emphasize that the fits considered here are
primarily included only for comparison and to provide
an alternative method for approximating the histogram
moments.  They are not intended to be unique descriptors of the true form
of the underlying pdfs. On a related note, we also stress that the fitted
moments listed in Table 1 are only approximate and are sensitive to the
 form of the fitting function.  For example, the kurtosis of the L1228
pdf varies by 62\% depending on whether the tails are taken to be exponential
or Gaussian (Table 1).  The statistics in the far tails, which give a
significant contribution to the high-order moments, are simply not good
enough in most cases to make a unique, unambiguous identification and
extrapolation, even if the pdfs were truly well described by simple
functional forms.  Also, the uncertainty in each centroid velocity
arising from instrumental noise can distort the pdfs, leading to an
increase in the effective histogram bin size due to the ``spillover''
from neighboring bins.  A complete treatment of this effect will be
deferred to a later paper (paper II), but for now we note that the
influence of this ``spillover'' on the values of the skewness and kurtosis
has been estimated and is found to be typically less than 20\%
(but possibly up to 50\%, with the largest discrepancies occurring in L1228
and Ori B, region 2).  It should also be pointed out that other
methods of pdf estimation, such as kernel and
parametric estimators (Vio {\em et al.} 1994), may be preferable to the
classical histogram estimator used here.

Despite these difficulties and qualifications, it is evident that many of the
pdfs shown
in Figure 1 exhibit broad, often nearly exponential, tails indicative of
highly non-Gaussian fluctuating velocity structure.
In particular, there is good evidence for nearly exponential behavior
in the tails of all of the composite spectra, along with
L1551, Ori B regions 3 and 4, and possibly L1228 and
Ori B region 1a, although the latter is questionable (recall that this
region, Ori B 1a, is the smallest of the data sets, composed of
only several hundred independent spectra).  Secondary tail components
with a steeper behavior, approaching Gaussian, may be present in HH83,
Ori B region 2, and possibly L1228 and Ori B region 1c, although
again, the latter has both the largest discrepancy and the
poorest statistics.  Note that the sharp cutoff at $v\sim 2.2$\kms in
Ori B, region 3 (Fig. 1k), and in the Ori B and all regions composites
(Figs. 1n and o), which are dominated by this region at high
velocities, are likely an artifact of the manner in which the centroid
velocities for this region were computed.
As mentioned above, region 3 overlaps regions 2 and 4 in
space but lies at a lower velocity, and in order to isolate the emission
from this region, it was necessary to sum over only part of the profile,
cutting off at high velocities to avoid contributions (and subsequent
biasing of the centroid velocity) from the other components.  Two upper
cutoffs were used for the calculation of the centroid velocity (see MB),
the first (used for 60\% of the points) was 9.2 \kms (lsr) and the
second (including the remaining 40\%) was 7.5 \kms (lsr).  Since the
horizontal axis of Fig. 1(k) is centered about an lsr velocity of 5.04
\kms, then it is likely that the high velocity cutoff in the pdf
at 2.2 \kms (which translates to about 7.2\kms lsr) is due to the high
velocity cutoff in the calculation of the centroids and not to any sharp
variation in the true pdf.  We have taken this cutoff into account when
calculating the skewness of Ori B, region 3, and the Ori B and ``all
regions'' composites by excluding all points with velocities
$|v| \gsim 2.2$\kms, which included the anomalously low point at
the high velocity end and several (2 to 4) points on the low velocity
end, so that the pdfs would not be biased toward low
velocities.  The remaining entries in Table 1 for Ori B, region 3,
however, were calculated using the full histogram.

The often negative (but fairly well distributed)
skewnesses of the
centroid pdfs are in contrast to  the positive values typically seen in the
average line profiles for
each region, which, particularly in the cases of HH83, L1551, and L1228,
show some excess blue-shifted emission likely due to
molecular outflows whose red-shifted lobes are
partially shrouded or otherwise absent.
This suggests again that the line-of-sight averaging and the
space filtering inherent in the centroid pdf approach make it less sensitive
to systematic motions (such as outflow bubbles) than are individual or
average profiles for the same region.

It is interesting to compare the pdfs of Figure 1 with the
line profiles observed by Falgarone \& Phillips (1990),
which sample radial velocity fluctuations along the line of sight
and which exhibit relatively broad wings.  In order to quantify
the excess wing emission, they fit their observed profiles to two
separate Gaussian components.
Falgarone and Phillips were primarily interested in demonstrating
the similar core/wing scaling on
very different scales, and so were careful not to draw conclusions
about
the shapes of the wings, although they remark that most of the profiles
were better fit with two Gaussians, and in only a few cases were better
fit with exponential or Lorentzian wings.
For the  two-Gaussian fits, they find typical dispersion ratios
(wing/core, hereafter $D$) of $3.3\pm0.2$ and amplitude
ratios (wing/core, hereafter $P$) ranging between about $0.03$ and
$0.8$.
The sum of two centered Gaussians with these values for $P$ and $D$
give kurtoses between 3.9 and 10.2,
a range that is higher than most of the values listed in Table 1.
Of the pdfs in Figure 1 which show an approximately Gaussian secondary
component, the corresponding ratios are:  $D=1.45$ and $P=0.44$ for
HH83, $D=2.5$ and $P=0.067$ for L1228, $D=2.1$ and $P=0.063$ for
Ori B region 1c, and $D=1.9$ and $P=0.097$ for the two strongest
components in Ori B region 2.
The somewhat smaller kurtoses and dispersion ratios in the present work
(despite the much less quiescent nature of
the regions we have studied) may in part be a result
of the line-of-sight averaging inherent in the centroid pdf approach,
which will tend to make the distribution more Gaussian if the velocity
structure is indeed stochastic (this follows from the central limit
theorem).  It is also possible in principle that the larger values
of $D$ found for the individual profiles could be due in part to a
warmer, more rarefied molecular ``inter-clump'' medium, which
would be apparent in the profiles (provided the column density is
sufficient), but, due to its symmetry,
would not show up in the centroid velocity measurements.
Thus, the core fraction in the centroid pdfs (Table 1) relative to that
observed in individual or average profiles can be used to set column
density limits on such a warm, molecular medium, if the cloud and
its surrounding medium are optically thin.  Also, as mentioned
above, ``optical depth broadening'' could give rise to excess wing
emission that would have no counterpart in the centroid pdfs.
However, as noted by Falgarone and Phillips (1990),
the line shapes, at least for their data, seem to rule out
optical depth effects, and
the variation among the wing dispersions make it
unlikely that warm gas is a dominant component.

\section{Comparison with Theory}

        A number of theoretical models provide mechanisms to produce
broad tail components in the velocity pdfs,
although none of them explain why the
tails are nearly exponential in some regions, as found here.
Collisions of clouds (Keto and Lattanzio 1989) or
gas oscillations and streams connected with clumps which might
represent non-linear waves or other phenomena
(Elmegreen 1990, 1994) produce a ``core-envelope'' density structure with
distinct velocity distributions.
Greaves and White (1991) have interpreted a velocity
discontinuity in the OMC1 ridge as evidence for such a collision.
Elmegreen's model assumes that the gas is composed of unresolved clumps with
intrinsic velocity distributions in both
the core and envelope regions that are Gaussian
from locally randomized motions and clump-clump interactions
according to the central limit theorem,
with superimposed streaming motions from non-linear waves.
By further assuming that the streaming part has
a power law velocity-density scaling similar to that
expected for non-linear MHD waves
and related processes (Adams and Fatuzzo 1993),
Elmegreen (1990) showed that the resulting composite velocity
distribution, integrated over the density gradient along the line of flow
symmetry and {\em inside} a single beam or averaged over over many beams,
agrees well with the broad wing profiles observed by Falgarone
and Phillips (1990),
including the universal 3:1 wing:core dispersion ratio obtained from a
two-Gaussian fit.
Elmegreen (1994) discussed how the relative strengths
of the core and wing might depend on time. The kurtoses of these pdfs
(Elmegreen, personal
communication) span the range found in the present work, varying with the
assumed density contrast and relative amplitudes of the core and envelope.
However the correspondence between this integral average along the density
gradient and the spatially sampled centroid pdf investigated here is
unclear -- primarily
because the pdf of the model depends on how well the streaming motions
and clump-clump dispersions are resolved.
The spatial pdf of {\em resolved} clumps and flows
will depend on spatial orientation and other parameters; presumably a
smooth and symmetric centroid-pdf comparable to those observed here
will result from an average over a
large number of lines of sight, if the scale of these motions is still
much smaller than the scale of the sampled region.
If the clumps are {\em unresolved}, then the envelopes will
only contribute to the average spectral line wings (which will be broad)
but not to the centroid velocity {\em within}
a single beam; this will produce
a centroid pdf that is just determined by the motions of the core regions.
The best agreement between the models and the observations arises if
the colliding flows (or some of them) are resolved;
then the velocity distribution for all of the pieces of the flow
will produce broad tails on the
centroid pdf instead of broad linewings on the individual spectra.
With only partial resolution, both the centroid pdf and the
individual spectra will have broad tails or wings.

         The structures found by Keto and Lattanzio (1989) are the result of
3-dimensional hydrodynamic simulations of cloud collisions. Their Figures 4-6
are images of the velocity centroids that are directly comparable with the
observed centroid images used here, although the pdfs were not presented.
(Keto and Lattanzio {\em do} give a montage of line profiles along individual
lines of sight, which show a great variety of shapes.)
However, because the flows are not assumed to be unresolved, and
because the wing emission is due to the streaming motion of the envelope
(rather
than the dispersion of the clumps posited by Elmegreen), the velocity field
will depend on initial conditions, time, and especially viewing angle.  An
average over these parameters would be necessary to compare with the
densely sampled regions studied here, which are likely to include a number
of such single-event collisions.  It might be possible to obtain flat
(relative to a Gaussian) wings in the cloud collision model if the
inflowing material has, say, a power-law velocity profile as a function of
distance from the compressed layer, but we have insufficient data from the
simulations to check this.  Hopefully future simulations will provide the
necessary information.

        Rather than attribute the pdf tails to such highly compressible
collisions or streams, another
possibility is that the pdf tails are due to an excess of high-velocity
events evoked by the vortical part of the advection operator, which causes
the velocity derivative and dissipation field to become highly localized
(e.g. in vorticity tubes in 3D incompressible turbulence).  This process,
often referred to as ``intermittency'' because a stationary probe in the flow
records a ``bursty'' velocity time series as the high-intensity patches flow
by, is the explanation favored by Falgarone and Phillips, although they
were referring to the pdf of velocity differences or derivatives, not the
velocity itself (we note again the problem with interpreting the line
profile as a pdf of velocity {\em differences}). However it is known
from turbulence simulations that even purely incompressible nonlinear
advection (in concert with pressure gradients) can produce broad velocity
pdf tails at small enough scales, especially in the dissipation range
(e.g. Yamamoto and Kambe 1991, Fig.1)
and at not-so-small scales in convective MHD
turbulence at low Mach numbers (Brandenburg et al. 1994), although the
latter may include other effects arising from magnetic fields and
rotation.
These studies suggest that even in the compressible,
self-gravitating case nonlinear advection may play
a major role.  In this regard the low-resolution (20x20)
2-dimensional simulations of rotating self-gravitating, compressible
turbulence (a magnetic field was {\em only} included by taking a polytropic
index of 2) by Yue et al. (1993) are of interest.  These calculations give
broad wings similar to the observations of Falgarone and Phillips; given
the low resolution, it is unlikely that cloud or stream collisions could
play a role, so perhaps nonlinear advection coupled with gravity (and
possibly rotation) is all that is required.
The advective effects cannot be due to the usual
association of intermittency with vorticity stretching, though, since this
term does not exist in two dimensions.
The very low resolution is of concern, however, as
well as the possible dependence on the imposed symmetry assumptions and
unusual initial conditions.   Again there is no indication of how these
simulations can account for the nearly exponential wings or the
{\em variations}  in the observed pdfs (e.g. some are nearly Gaussian).

        Concerning the importance of magnetic effects in contributing
to
broad pdf tails, we know of only two relevant results.
First,
the simulation of hydrodynamic particles with imposed wave forcing
meant to
represent magnetic fields by Stenholm and Pudritz (1993), do {\em not}
show any excess wings in line profiles, although it may be argued that
this is not
really an MHD
simulation, since the waves are simply imposed on the gas with some
frequency and amplitude.  On the other hand, the simulations
of 3-D, rotating, thermally forced MHD turbulence
by Brandenburg et al. (1994) do exhibit broad tails in the
{\em velocity} pdfs, at least in the horizontal directions
(the pdf of vertical velocity is dominated by
convective motions such as updrafts and plumes and is therefore not as
directly comparable to the present results, which are for velocity
{\em fluctuations}).

        Higher-resolution ($512^3$) 3-dimensional simulations of
decaying compressible turbulence (without self-gravity or magnetic
fields) have been performed
by Porter, Pouquet, and Woodward (1993), and Falgarone et al. (1994) present
centroid velocity pdfs for individual 32x32 subregions.  This is just the
type of analysis that is needed for a comparison with line shapes or pdf
forms.  The pdfs of the subregions show a wide variation in appearance,
and, in particular, kurtosis (Lis, personal communication), but these are
dominated by coherent motions on the scale of the subregions or larger.
When the entire simulation box is included, the velocity pdf is nearly
Gaussian.  These simulations may be applicable to some quiescent clouds,
but since they do not include self-gravity or internal stellar sources of
energy input, they are not directly comparable to the observations of
active regions presented here.

       We expect that the influence of internal star formation is crucial
in understanding the observed pdfs of active regions.  Two-dimensional
simulations including heating or stirring by star formation and cooling
have been given by several groups (see Bania and Lyon 1980, Balser, Bania,
and Lyon 1990, Rosen, Bregman, and Norman 1993, Vazquez-Semadeni, Passot,
and Pouquet 1994).  These simulations variously include self-gravity
(Vazquez-Semadeni et al.), radiative transfer and ionization (Bania et al.
group), and the stellar component as a fluid (Rosen et al.), but the
important point in common is the inclusion of energy and momentum input due
to young massive stars, whose formation rate is parameterized in terms of
some assumed dependence on the density and perhaps other variables.
Unfortunately, a velocity pdf analysis of the type provided by Falgarone et
al. (1994) is unavailable for these simulations.  The present work suggests
that the ability of these models to yield a range of velocity pdfs from
nearly Gaussian to Gaussian cores with nearly exponential tails, will
provide a critical test of the models, although it must be again
remembered that in two dimensions vorticity stretching is absent; the
importance of vorticity stretching might be implicated by the absence of
exponential wings in 2-dimensional simulations that include much of the
otherwise important physics (Vazquez-Semadini, personal communication).
Preliminary work on the velocity pdfs of the simulations of
Vazquez-Semadini et al. is underway.

        Although collisions, nonlinear advection, and star formation
activity all suggest themselves as explanations of the broad tails, it
is clear after some thought that in an active star-forming region these
three processes cannot be separated; for example, stellar winds and
ionization drive flows that cause collisions which may form more stars,
etc.   For this reason it appears that careful analysis of numerical
simulations afford the best theoretical approach to understanding the
observed phenomena of broad pdf velocity tails.  In the meantime, we can
find no theoretical model that explains why the tails would be nearly
exponential, Perhaps a mapping model of the advection term similar to that
discussed by, e.g. She et al. (1993), but generalized to the compressible
or even magnetized case will be useful.

In summary, we have investigated the fluctuating velocity structure in
several nearby molecular clouds using the probability distribution
function of emission line centroid velocities and we find evidence for
non-Gaussian behavior in most of the regions we have studied.  The most
important results seem to be the variation between subregions, and the nearly
exponential tails found in several cases.  Although several theoretical
interpretations have been discussed, in most cases the theoretical data is
insufficient for a comparison with these observations, although we have found
no explanation for the nearly exponential tails.

 \acknowledgements
 We would like to thank John Bally again for providing the data sets used
 in this work and for countless discussions regarding their analysis and
 interpretation.  Bruce Elmegreen, Darek Lis and Enrique Vazquez-Semadini
 kindly provided data on
the pdf moments for their models.  We thank Edith Falgarone and Bruce
Elmegreen for comments on an earlier version of this manuscript.
MSM would also like to thank Ellen Zweibel and Juri Toomre
for their insight, encouragement, and financial support.

\newpage
\parskip=8mm
\parindent=4mm

\end{document}